\documentclass[aps,prc,twocolumn,showpacs,superscriptaddress,amsmath,amssymb]{revtex4}
\usepackage{epic}
\usepackage{curves}
\usepackage{graphicx}

\begin{document}


\title{First direct lifetime measurement of the $2^+_1$ state in $^{72,74}$Zn:\\
new evidence for shape transition between $N=40$ and $N=42$ close to $Z=28$  }


\author{M.~Niikura}
\email[]{niikura@ipno.in2p3.fr}
\affiliation{Institut de Physique Nucl\'eaire (IPN), IN2P3-CNRS, Universit\'e Paris-Sud 11, F-91406 Orsay Cedex, France}
\author{B.~Mouginot}
\affiliation{Institut de Physique Nucl\'eaire (IPN), IN2P3-CNRS, Universit\'e Paris-Sud 11, F-91406 Orsay Cedex, France}
\author{S.~Franchoo}
\affiliation{Institut de Physique Nucl\'eaire (IPN), IN2P3-CNRS, Universit\'e Paris-Sud 11, F-91406 Orsay Cedex, France}
\author{I.~Matea}
\affiliation{Institut de Physique Nucl\'eaire (IPN), IN2P3-CNRS, Universit\'e Paris-Sud 11, F-91406 Orsay Cedex, France}
\author{I.~Stefan}
\affiliation{Institut de Physique Nucl\'eaire (IPN), IN2P3-CNRS, Universit\'e Paris-Sud 11, F-91406 Orsay Cedex, France}
\author{D.~Verney}
\affiliation{Institut de Physique Nucl\'eaire (IPN), IN2P3-CNRS, Universit\'e Paris-Sud 11, F-91406 Orsay Cedex, France}
\author{F.~Azaiez}
\affiliation{Institut de Physique Nucl\'eaire (IPN), IN2P3-CNRS, Universit\'e Paris-Sud 11, F-91406 Orsay Cedex, France}
\author{M.~Assie}
\affiliation{Institut de Physique Nucl\'eaire (IPN), IN2P3-CNRS, Universit\'e Paris-Sud 11, F-91406 Orsay Cedex, France}
\affiliation{Grand Acc\'el\'erateur National d'Ions Lourds (GANIL), CEA/DSM-CNRS/IN2P3, BP 55027, F-14076 Caen Cedex 05, France}
\author{P.~Bednarczyk}
\affiliation{The Henryk Niewodnicza\'nski Institute of Nuclear Physics (IFJ PAN), PL-31342 Krak\'ow, Poland}
\author{C.~Borcea}
\affiliation{Horia Hulubei National Institute for Physics and Nuclear Engineering (IFIN-HH), P.O. Box MG-6, ROM-76900 Bucharest-Magurele, Romania}
\author{A.~Burger}
\affiliation{Department of Physics, University of Oslo, N-0315 Oslo, Norway}
\author{G.~Burgunder}
\affiliation{Grand Acc\'el\'erateur National d'Ions Lourds (GANIL), CEA/DSM-CNRS/IN2P3, BP 55027, F-14076 Caen Cedex 05, France}
\author{A.~Buta}
\affiliation{Horia Hulubei National Institute for Physics and Nuclear Engineering (IFIN-HH), P.O. Box MG-6, ROM-76900 Bucharest-Magurele, Romania} 
\author{L.~C\'aceres}
\affiliation{Grand Acc\'el\'erateur National d'Ions Lourds (GANIL), CEA/DSM-CNRS/IN2P3, BP 55027, F-14076 Caen Cedex 05, France} 
\author{E.~Cl\'ement}
\affiliation{Grand Acc\'el\'erateur National d'Ions Lourds (GANIL), CEA/DSM-CNRS/IN2P3, BP 55027, F-14076 Caen Cedex 05, France}
\author{L.~Coquard}
\affiliation{Institut f\"ur Kernphysik, der Technische Universit\"at Darmstadt, D-64289 Darmstadt, Germany}
\author{G.~de~Angelis}
\affiliation{Laboratori Nazionali di Legnaro, INFN, I-35020 Legnaro, Padova, Italy}
\author{G.~de~France}
\affiliation{Grand Acc\'el\'erateur National d'Ions Lourds (GANIL), CEA/DSM-CNRS/IN2P3, BP 55027, F-14076 Caen Cedex 05, France}
\author{F.~de~Oliveira~Santos}
\affiliation{Grand Acc\'el\'erateur National d'Ions Lourds (GANIL), CEA/DSM-CNRS/IN2P3, BP 55027, F-14076 Caen Cedex 05, France}
\author{A.~Dewald}
\affiliation{Institut f\"ur Kernphysik, Universit\"at zu K\"oln, Z\"ulpicher Stra{\ss}e 77, D-50937 K\"oln, Germany}
\author{A.~Dijon}
\affiliation{Grand Acc\'el\'erateur National d'Ions Lourds (GANIL), CEA/DSM-CNRS/IN2P3, BP 55027, F-14076 Caen Cedex 05, France}
\author{Z.~Dombradi}
\affiliation{Institute of Nuclear Research of the Hungarian Academy of Sciences (ATOMKI), H-4001 Debrecen, P.O. Box 51, Hungary}
\author{E.~Fiori}
\affiliation{CSNSM, UMR 8609, IN2P3-CNRS, Universit\'e Paris-Sud 11, F-91405 Orsay Cedex, France}
\author{C.~Fransen}
\affiliation{Institut f\"ur Kernphysik, Universit\"at zu K\"oln, Z\"ulpicher Stra{\ss}e 77, D-50937 K\"oln, Germany}
\author{G.~Friessner}
\affiliation{Institut f\"ur Kernphysik, Universit\"at zu K\"oln, Z\"ulpicher Stra{\ss}e 77, D-50937 K\"oln, Germany}
\author{L.~Gaudefroy}
\affiliation{CEA, DAM, DIF, F-91297 Arpajon, France}
\author{G.~Georgiev}
\affiliation{CSNSM, UMR 8609, IN2P3-CNRS, Universit\'e Paris-Sud 11, F-91405 Orsay Cedex, France}
\author{S.~Gr\'evy}
\affiliation{Grand Acc\'el\'erateur National d'Ions Lourds (GANIL), CEA/DSM-CNRS/IN2P3, BP 55027, F-14076 Caen Cedex 05, France}
\author{M.~Hackstein}
\affiliation{Institut f\"ur Kernphysik, Universit\"at zu K\"oln, Z\"ulpicher Stra{\ss}e 77, D-50937 K\"oln, Germany}
\author{M.~N.~Harakeh}
\affiliation{Grand Acc\'el\'erateur National d'Ions Lourds (GANIL), CEA/DSM-CNRS/IN2P3, BP 55027, F-14076 Caen Cedex 05, France}
\affiliation{Kernfysisch Versneller Instituut (KVI), University of Groningen, NL-9747 AA Groningen, The Netherlands}
\author{F.~Ibrahim}
\affiliation{Institut de Physique Nucl\'eaire (IPN), IN2P3-CNRS, Universit\'e Paris-Sud 11, F-91406 Orsay Cedex, France}
\author{O.~Kamalou}
\affiliation{Grand Acc\'el\'erateur National d'Ions Lourds (GANIL), CEA/DSM-CNRS/IN2P3, BP 55027, F-14076 Caen Cedex 05, France}
\author{M.~Kmiecik}
\affiliation{The Henryk Niewodnicza\'nski Institute of Nuclear Physics (IFJ PAN), PL-31342 Krak\'ow, Poland}
\author{R.~Lozeva}
\affiliation{CSNSM, UMR 8609, IN2P3-CNRS, Universit\'e Paris-Sud 11, F-91405 Orsay Cedex, France}
\affiliation{IPHC, IN2P3-CNRS, Universit\'e de Strasbourg, 67037 Strasbourg, France}
\author{A.~Maj}
\affiliation{The Henryk Niewodnicza\'nski Institute of Nuclear Physics (IFJ PAN), PL-31342 Krak\'ow, Poland}
\author{C.~Mihai}
\affiliation{Horia Hulubei National Institute for Physics and Nuclear Engineering (IFIN-HH), P.O. Box MG-6, ROM-76900 Bucharest-Magurele, Romania}
\author{O.~M\"oller}
\affiliation{Institut f\"ur Kernphysik, der Technische Universit\"at Darmstadt, D-64289 Darmstadt, Germany}
\author{S.~Myalski}
\affiliation{The Henryk Niewodnicza\'nski Institute of Nuclear Physics (IFJ PAN), PL-31342 Krak\'ow, Poland}
\author{F.~Negoita}
\affiliation{Horia Hulubei National Institute for Physics and Nuclear Engineering (IFIN-HH), P.O. Box MG-6, ROM-76900 Bucharest-Magurele, Romania}
\author{D.~Pantelica}
\affiliation{Horia Hulubei National Institute for Physics and Nuclear Engineering (IFIN-HH), P.O. Box MG-6, ROM-76900 Bucharest-Magurele, Romania}
\author{L.~Perrot}
\affiliation{Institut de Physique Nucl\'eaire (IPN), IN2P3-CNRS, Universit\'e Paris-Sud 11, F-91406 Orsay Cedex, France}
\author{Th.~Pissulla}
\affiliation{Institut f\"ur Kernphysik, Universit\"at zu K\"oln, Z\"ulpicher Stra{\ss}e 77, D-50937 K\"oln, Germany}
\author{F.~Rotaru}
\affiliation{Horia Hulubei National Institute for Physics and Nuclear Engineering (IFIN-HH), P.O. Box MG-6, ROM-76900 Bucharest-Magurele, Romania}
\author{W.~Rother}
\affiliation{Institut f\"ur Kernphysik, Universit\"at zu K\"oln, Z\"ulpicher Stra{\ss}e 77, D-50937 K\"oln, Germany}
\author{J.~A.~Scarpaci}
\affiliation{Institut de Physique Nucl\'eaire (IPN), IN2P3-CNRS, Universit\'e Paris-Sud 11, F-91406 Orsay Cedex, France}
\author{C.~Stodel}
\affiliation{Grand Acc\'el\'erateur National d'Ions Lourds (GANIL), CEA/DSM-CNRS/IN2P3, BP 55027, F-14076 Caen Cedex 05, France}
\author{J.~C.~Thomas}
\affiliation{Grand Acc\'el\'erateur National d'Ions Lourds (GANIL), CEA/DSM-CNRS/IN2P3, BP 55027, F-14076 Caen Cedex 05, France}
\author{P.~Ujic}
\affiliation{Grand Acc\'el\'erateur National d'Ions Lourds (GANIL), CEA/DSM-CNRS/IN2P3, BP 55027, F-14076 Caen Cedex 05, France}
\affiliation{Vinca Institute of Nuclear Sciences, University of Belgrade, Nos. 12--14 Mike Alasa, 11001 Vinca, Serbia}


\date{\today}

\begin{abstract}
  We report here the first direct lifetime measurement of the
  2$^{+}_{1}$ state in $^{72,74}$Zn. The neutron-rich beam was
  produced by in-flight fragmentation of $^{76}$Ge at the Grand
  Acc\'el\'erateur National d'Ions Lourds and separated with the LISE
  spectrometer. The 2$^{+}_{1}$ state was excited by inelastic
  scattering and knock-out reaction on a CD$_2$ target and its
  lifetime was measured by the recoil-distance Doppler-shift method
  with the K\"oln plunger device combined with the EXOGAM
  detectors. The lifetimes of the 2$^{+}_{1}$ states in $^{72,74}$Zn
  were determined to be 17.9(18) and 27.0(24)~ps, which correspond to
  reduced transition probabilities $B(E2;2^{+}_{1}\rightarrow
  0^{+})=385(39)$ and 370(33)~$e^2\textrm{fm}^4$, respectively.
  These values support the idea of a systematic maximum of
  collectivity at $N=42$ for Zn, Ge and Se nuclei. In addition, the
  avalable systematics in the neighboring nuclei point towards a
  transition from a spherical oscillator at $N=40$ to complete
  $\gamma$-softness at $N=42$.
\end{abstract}


\maketitle

\begin{figure*}
  \centering
  \includegraphics[width=.9\textwidth]{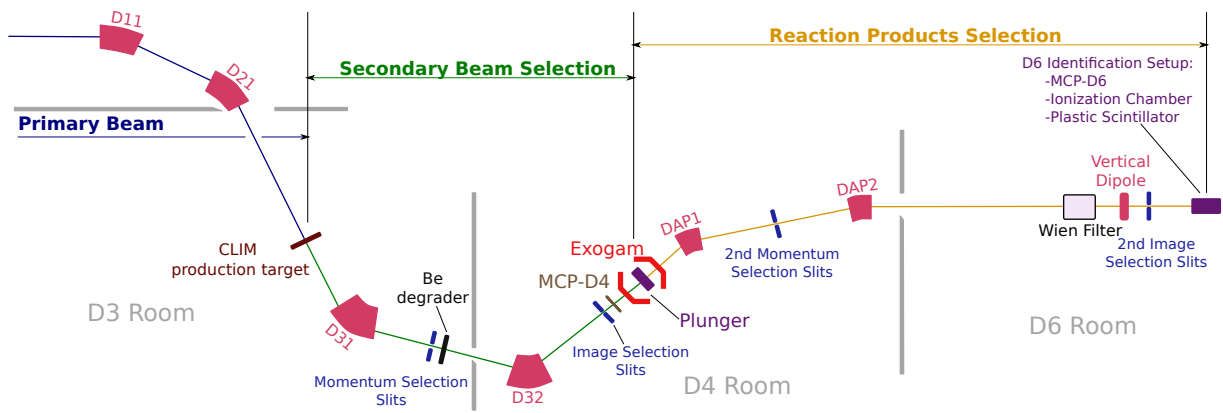}
  \caption{\label{fig_setup} (Color online) Schematic view of the
    experimental setup. In this measurement LISE was used both for
    secondary-beam selection (rooms D3-D4) and for reaction-products
    selection from secondary target (rooms D4-D6). The secondary beam
    was produced in the CLIM production target~\cite{clim1,clim2}
    installed in D3. The K\"oln plunger device~\cite{plunger} was
    placed in the first achromatic focal point (D4), surrounded by
    eight EXOGAM clovers. $\Delta$E-E telescope consisting of an
    ionization chamber and a plastic scintillator was placed in D6.}
\end{figure*}

\section{Introduction}
There has been a continuous interest in the study of the structure
effects associated with $N=40$ close to $Z=28$ since the original
publication by Bernas~\textit{et al.}~\cite{ber} and the magic
character of $^{68}_{28}$Ni$_{40}$ has been extensively discussed (for
instance see Ref.~\cite{lan} and references therein). One of the most
interesting features in the immediate vicinity of the $Z=28$, $N=40$
crossing was pointed out by Perru~\textit{et al.}~\cite{per} and
consists in an apparent over occupation of the neutron $1g_{9/2}$
orbit, starting already for $N\le40$, and a concomitant strong proton
core polarization effect. These phenomena were invoked to explain the
sudden and large increase of a reduced transition probability $B(E2)$
observed between $^{68}$Ni$_{40}$ and $^{70}$Ni$_{42}$. Independently,
there has been considerable accumulation of experimental evidences
pointing toward a sudden structural change from $N=40$ to $42$ in the
Ge chain. This change was originally inferred from the analysis of
nucleon transfer cross sections~\cite{ver} and was seen to be also
present in the Se chain~\cite{rot} though somewhat attenuated. This
phenomenon is also clear when considering $B(E2)$ and $Q(2^+_1)$
data~\cite{lec}. It was soon admitted that the Ge and Se nuclei
undergo a maximum of collectivity (in the most general sense) at
$N=42$~\cite{nol,ver2,lec} and that a new collectivity regime (the
nature of which will be discussed in last section) develops at this
particular neutron number to the end of the neutron $1g_{9/2}$
shell. From the comparison with Ge and Se systematics one could
naturally expect a maximum of collectivity at $N=42$, or more
generally structural change from $N=40$ to $42$, also in the Zn
chain. In addition, the microscopic mechanism that would connect the
strong polarization effect affecting the $B(E2)$ value in
$^{70}$Ni$_{42}$ to the $N=42$ collectivity maximum in Ge and Se is
still difficult to establish in a direct way. The over occupation of
the neutron $1g_{9/2}$ orbit for $Z\ge30$, already hinted at in
Ref.~\cite{per}, is confirmed by recent shell-model
calculations~\cite{hon}: this can be seen, for instance, in Fig.~24 of
Ref.~\cite{hon}, which shows calculated occupation of the neutron
$1g_{9/2}$ orbit in the $0^+_g$ and $2^+_g$ of the even-even Ge
chain. However, in these calculations the proton $1f_{7/2}$ orbit is
not included in the valence space and the effect of the proton core
polarization cannot be quantified directly \textit{i.e.} if such an
effect is indeed accounted for, it is only through a very indirect
way, that is \textit{via} some modification of the residual
interaction after fitting some of the two-body matrix elements to data
on nuclei close to $Z=28$. More recently shell-model calculations in a
larger valence space including the $fp$ shell for protons and the
$1f_{5/2}$, $2p_{3/2}$, $2p_{1/2}$, $1g_{9/2}$ and $2d_{5/2}$ orbitals
for neutrons became available~\cite{len}. The strong increase of
$B(E2)$ values from $^{68}$Ni$_{40}$ to $^{70}$Ni$_{42}$ is somewhat
reproduced using standard effective charges but not quite (see Fig.~5
in Ref.~\cite{len}). Interestingly, a second calculation in the same
valence space but with a slightly revised version of the residual
interaction worsens a little bit the result~\cite{sie}, as if the
inclusion of the $\Delta \ell=2$ quadrupole partners
$g_{9/2}$-$d_{5/2}$, contrary to what had been expected earlier, was
not a sufficient ingredient to explain the apparent increase of
``collectivity'' in Ni between $N=40$ and $42$. Even-even Zn nuclei
close to $N=40$ should then provide the ideal test ground if one is to
try to pin down the potential connection between the two phenomena:
strong proton core polarization and maximum of collectivity at $N=42$.

In this paper we propose to address this question by means of a direct
lifetime measurement of the $2^+_1$ state in $^{72}$Zn$_{42}$ and
$^{74}$Zn$_{44}$. We expect in that way to get a more accurate
evaluation of the $B(E2)$ values than those obtained from the previous
scattering cross-section measurements \cite{lee,per,vdw}, which are
generally sensitive to the energy regime and/or model dependent in
some way. In particular, when not all the $E2$ strength connecting the
lowest-lying states can be observed \textemdash which is usually the
case in radioactive beam experiments \textemdash some value for the
quadrupole moment of the $2^+$ state must be assumed. Such kind of
assumptions is particularity hazardous in a region of transitional
nuclei characterized by subtle and complicated collective effects.

\section{Experimental condition}

The experiment was performed at the Grand Acc\'el\'erateur National
d'Ions Lourds (GANIL) using the recoil-distance Doppler-shift method
(RDDS) method~\cite{ddc} applied to the intermediate-energy reaction.
The LISE spectrometer~\cite{lise} was used both for the separation of
the reaction fragments in the first half of the spectrometer and for
the identification of the reaction recoils in the second half.
Details of the experimental setup are shown in Fig.~\ref{fig_setup}.

A cocktail beam of $^{73,74}$Zn and $^{72}$Cu was produced by the
projectile-fragmentation reaction of a 60-MeV/nucleon
$^{76}$Ge$^{30+}$ beam impinging on the CLIM target~\cite{clim1,clim2}
consisting of a 580-$\mu$m-thick $^{9}$Be. A 500-$\mu$m-thick $^{9}$Be
wedge degrader was placed in the first dispersive focal plane of LISE
to select nuclei of interest produced in the target. A total average
intensity of 1.0$\times$10$^5$ particles per second containing 75\%
$^{74}$Zn was measured at D4 with 1.0~$e\mu$A of the primary beam
intensity.

The K\"oln plunger device~\cite{plunger} used for the RDDS measurement
was mounted with a 445-$\mu$m-thick (35.5-mg/cm$^2$) CD$_2$ target and
a 273-$\mu$m-thick (50.5-mg/cm$^2$) $^9$Be degrader at the first
achromatic focal point of LISE (D4). The distance between the target
and the degrader can be varied by changing the degrader position with
a relative precision of 5~$\mu$m.  In the present experiment ten
different target-degrader distances ($d$) were set over the range from
0 to 20 mm as summarized in Table~\ref{tab_stat}.  The plunger device
was surrounded by eight segmented EXOGAM Ge
clovers~\cite{exogam1,exogam2,exogam3}, covering angles between
30$^{\circ}$ and 58$^{\circ}$ and between 117$^{\circ}$ and
148$^{\circ}$ with respect to the beam direction. The electronic
segmentation of the EXOGAM clovers reduces the Doppler broadening by
about 40\%, leading to an energy resolution (FWHM) of about 15~keV for
the 605.9-keV $2^{+}_{1} \rightarrow 0^{+}$ transition in $^{74}$Zn.

The identification of the nuclei was carried out by means of energy
loss and time-of-flight ($\Delta$E-TOF) measurements.  The $\Delta$E
was measured with an ionization chamber (CHIO) placed at the final
achromatic focal point of LISE (D6) in front of a plastic
scintillator. The TOF was measured between two microchannel plate
detectors (MCP)~\cite{galotte}: the first one was placed 1.107~m
upstream of the plunger device, and the second one installed in front
of the ionization chamber.  They allowed an accurate TOF measurement
for the recoil fragments between D4 and D6 (distance = 24.739~m).  The
efficiencies of MCPs were about 90\% and 95\% for D4 and D6,
respectively. The typical count rate in the D6 detectors was
$1.0\times 10^4$ counts per second.

The acquisition was triggered mainly by the coincidence between the
plastic scintillator in D6 and one or more EXOGAM detectors. The
coincidence window was set to 200~ns. About $3\times10^7$ triggered
events were corrected for each target-degrader distance settings as
summarized in Table~\ref{tab_stat}.

\begin{table}[tbp]
  \centering
  \caption{\label{tab_stat} Summary of total statistics accumulated
    and measured decay curve ($I_a'/(I_b'+I_a')$) in each of the ten
    target-degrader distances ($d$). Distances presented here were
    measured between the back side of the target and the front side of
    the degrader.}
  \begin{tabular}[c]{cccc}
    \hline\hline
    $d$ (mm) & Statistics & \multicolumn{2}{c}{$I_{\mathrm{a}}'/(I_{\mathrm{b}}'+I_{\mathrm{a}}')$}\\
    & & $^{72}$Zn & $^{74}$Zn \\
    \hline
    0.00 & 27114810 & - & -\\
    0.75 & 31411531 & 0.895(20) & 0.914(25)\\
    1.25 & 34776072 & 0.758(35) & 0.858(13)\\
    1.75 & 37650948 & 0.699(16) & 0.757(13)\\
    2.50 & 44109130 & 0.569(23) & 0.695(10)\\
    3.50 & 30070949 & 0.597(26) & 0.621(11)\\
    5.00 & 38184414 & 0.489(20) & 0.564(13)\\
    8.00 & 31748616 & 0.510(25) & 0.509(11)\\
    15.0 & 18374052 & 0.524(59) & 0.499(20)\\
    20.0 & 25005851 & 0.475(31) & 0.509(22)\\
    \hline\hline
  \end{tabular}
\end{table}

\section{analysis procedure and result}

\begin{figure*}
  \begin{center}
    \includegraphics[width=.92\textwidth]{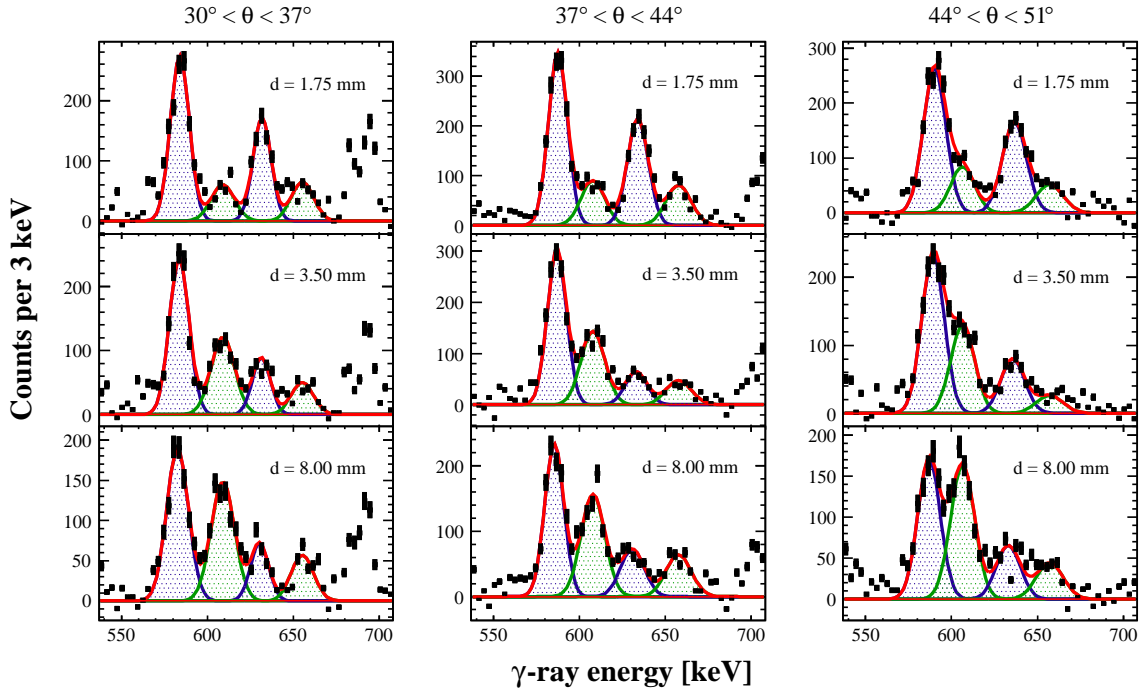}
    \caption{\label{fig_gamma} (Color online) Measured $\gamma$-ray
      spectra after Doppler-correction and background subtraction.
      They are gated on the prompt $\gamma$ timing within 10~ns. The
      four photopeaks observed in the spectra are associated with the
      $2^{+}_{1}\rightarrow 0^{+}$ transitions of $^{74,72}$Zn
      corresponding to the different recoil velocities before and
      after degrader, respectively. Solid lines represent the results
      of fitting with the Gaussian functions parameterized by GEANT4
      simulation. Gamma peaks shown in the range 680--700 keV at small
      angles are contaminants from $^{72}\mathrm{Ge}(n,n')$ (at
      834~keV in laboratory coordinate system.)}
  \end{center}
\end{figure*}

\begin{figure}
  \begin{center}
    \includegraphics[width=.48\textwidth]{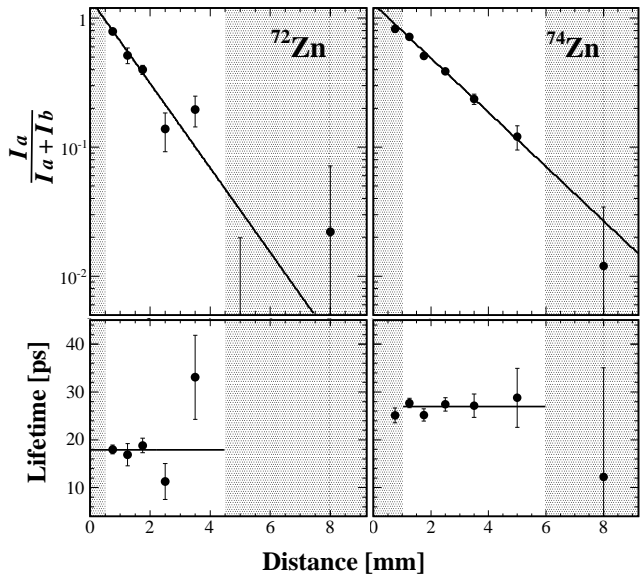}
    \vspace{-0.5cm}
    \caption{\label{fig_life} (Color online) Decay curves
      ($I_{\mathrm{a}}/(I_{\mathrm{b}}+I_{\mathrm{a}})$) (up) and
      lifetimes (down) of the 2$^{+}_{1}$ state in $^{72,74}$Zn (left
      and right) at several target-degrader distances. Decay curves
      are plotted in logarithmic scale. Points in the grayed areas are
      out of the ``region of sensitivity'' for the DDC
      method~\cite{ddc}.}
  \end{center}
\end{figure}

The lifetime of the $2^{+}_{1}$ state in $^{72,74}$Zn was deduced
based on the differential decay curve (DDC) method~\cite{ddc} applied
for intermediate energy reaction. To apply the DDC method, the decay
curve $I_{\mathrm{a}}/(I_{\mathrm{b}}+I_{\mathrm{a}})$ has to be
obtained from $\gamma$-ray spectra, where $I_{\mathrm{b}}$ and
$I_{\mathrm{a}}$ are the intensities of the $\gamma$ rays emitted
before and after the degrader from the 2$^{+}_{1}$ state of
$^{72,74}$Zn excited in the target.

In the present experiment, reactions in the degrader can also produce
the excited state of interest. Therefore, the extracted intensities
from the measured $\gamma$ spectra are not $I_{\mathrm{b}}$ and
$I_{\mathrm{a}}$, but $I_{\mathrm{b}}'$ and
$I_{\mathrm{a}}'$. $I_{\mathrm{b}}'$ represents the number of decays
from the $2^{+}_{1}$ state before the degrader, therefore
$I_{\mathrm{b}}=I_{\mathrm{b}}'$. On the other hand $I_{\mathrm{a}}'$
is the total number of decays after the degrader from the $2^{+}_{1}$
state. Therefore, $I_{\mathrm{a}}' = I_{\mathrm{a}} +
I_{\mathrm{deg}}$, where $I_{\mathrm{deg}}$ represents the decays from
the 2$^{+}$ state produced in the degrader. The decay curve can be
extracted as
\begin{equation}
  \frac{I_{\mathrm{a}}}{I_{\mathrm{b}}+I_{\mathrm{a}}} = (1+\alpha)\frac{I_{\mathrm{a}}'}{I_{\mathrm{b}}'+I_{\mathrm{a}}'} - \alpha,
  \label{eq_exp_decay2}
\end{equation}
with the production ratio between the degrader and the target: $\alpha
= I_{\mathrm{deg}}/(I_{\mathrm{b}}+I_{\mathrm{a}})$. In the case of a
sufficiently long distance between the target and the degrader, one
can assume $I_{\mathrm{a}} \simeq 0$ and $\alpha$ constant can be
deduced from eq.~(\ref{eq_exp_decay2}) as
\begin{equation}
  \alpha  = I_{\mathrm{a}}'/I_{\mathrm{b}}'.
  \label{eq_alpha}
\end{equation}

Figure~\ref{fig_gamma} shows the Doppler-corrected $\gamma$-ray
spectra measured for the three different emission angles of the
$\gamma$ rays~($\theta$) used in the Doppler correction and at three
out of ten target-degrader distances. For the lifetime determination
only forward EXOGAM detectors were used due to a 511~keV contamination
in the spectra of the backward detectors. From the forward-angle
detectors, segments with angles between 51$^{\circ}$ and 58$^{\circ}$
were removed from the analysis due to the poor separation between the
peaks of interest.

The geometrical efficiencies as a function of the $\gamma$-ray
emission point and the $\gamma$-ray energies as well as their widths
detected in EXOGAM were determined by a GEANT4
simulation~\cite{geant4-1,geant4-2}.  The simulation takes into
account the velocities of the incoming and outgoing particles, the
energy losses through the target and degrader, and the detector
geometry. Isotropic $\gamma$-ray emission with the Lorentz boost
effect and a given lifetime are assumed in the event generator of the
simulation. Therefore, a line-shape effect of the $\gamma$ peaks is
naturally included.

The intensities of the $\gamma$ rays emitted before and after the
degrader from the 2$^{+}_{1}$ state of $^{72,74}$Zn ($I_{\mathrm{b}}'$
and $I_{\mathrm{a}}'$) were extracted by fitting the region of
interest with four Gaussian functions as shown in
Fig.~\ref{fig_gamma}. The mean and sigma values for the Gaussian
functions are fixed from the simulation using mean velocities between
target and degrader ($\beta = v/c$) of 0.2449(11) and 0.2433(11) for
$^{72}$Zn and $^{74}$Zn, respectively. These $\beta$ values were
obtained from the alignment of Doppler-corrected peaks at all measured
angles. Obtained intensity ratios
$I_{\mathrm{a}}'/(I_{\mathrm{b}}'+I_{\mathrm{a}}')$, after taking into
account the geometrical efficiencies, are shown in
Table~\ref{tab_stat}.

The $\alpha$ constants were extracted by means of eq.~(\ref{eq_alpha})
and were found to be 1.00(7) and 1.01(6) for $^{72,74}$Zn,
respectively. The largest three distances (8.0, 15.0 and 20.0 mm) for
$^{72}$Zn and two (15.0 and 20.0 mm) for $^{74}$Zn were used for this
$\alpha$ constant extraction. Figure~\ref{fig_life} shows the result
of the DDC method applied to the decay curve obtained with
eq.~\ref{eq_exp_decay2}. The lifetimes were deduced to be 17.9(18) and
27.0(24)~ps for $^{72,74}$Zn, respectively. The errors are dominated
by the error on the $\alpha$ parameter (1.7 and 2.3~ps).

The side-feeding contribution to the lifetime from excitation of
higher energy states was examined. The upper limit of the 4$^{+}_{1}$
population was obtained to be 30\% of the 2$^{+}_{1}$ state population
from the intensity of the 847- ($^{72}$Zn) and 812-keV ($^{74}$Zn)
$4^{+}_{1}\rightarrow 2^{+}_{1}$ transition in the $\gamma$-ray
spectra. As long as the lifetime of the 4$^{+}_{1}$ state is estimated
to be less than 3~ps~\cite{vdw}, none of the data points at distances
in the sensitive region for the DDC method are affected by the side
feeding. No other $\gamma$ lines have been observed in
$\gamma$-$\gamma$ coincidence spectra gated on the 2$^{+}_{1}$ decay
of $^{72,74}$Zn.

\begin{figure}
  \begin{center}
    \includegraphics[width=.47\textwidth]{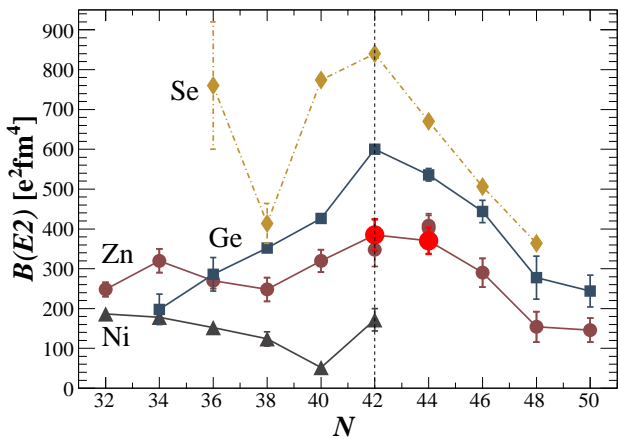}
    \caption{\label{be2sys} (Color online) Experimental systematics of
      $B(E2)$ for Ni (triangles), Zn (filled circles), Ge (squares)
      and Se (diamonds) isotopes. The values obtained within the
      present work are superimposed in red.}
  \end{center}
\end{figure}

The obtained lifetimes of 17.9(18) and 27.0(24)~ps correspond to
reduced transition probabilities $B(E2;2^{+}_{1}\rightarrow 0^{+}) =
385(39)$ and $370(33)$~$e^2\textrm{fm}^4$ for $^{72,74}$Zn,
respectively. These values are comparable within error bars to ones
extracted from Coulomb-excitation experiments with intermediate- and
low-energy $^{72,74}$Zn beams, which give $B(E2)$ values of
348(42)~$e^{2}$fm$^{4}$ (for $^{72}$Zn)~\cite{lee}, 408(30) and
401(32)~$e^{2}$fm$^{4}$ (for $^{74}$Zn)~\cite{per,vdw}. As can be seen
in Fig.~\ref{be2sys}, our value for $^{72}$Zn is closer to the higher
tip of the error bar, whereas for $^{74}$Zn it is closer to the lower
tip.
The $B(E2)$ systematics with new values obtained in the present work
shows a similar trend with that of Ge and Se nuclei, which have the
same local maximum at $N=42$. The consequence of this trend will be
discussed in the forthcoming section. In the previous work on
$^{74}$Zn~\cite{vdw}, the $B(E2)$ value was obtained using GOSIA code
on the assumption of the spectroscopic quadrupole moment
$Q(2^+_1)=0$~$e$b and the authors also provide a correlation between
the lifetime and $Q(2^+_1)$ (see Fig.~9 in Ref.~\cite{vdw}). From the
comparison between the present data and the previous Coulex
measurement a value of $Q(2^+_1)=+0.22$(30)~$e$b can be extracted for
$^{74}$Zn. The obtained $Q(2^+_1)$ of $^{74}$Zn will be also discussed
in the next section.

\section{Discussion}
\subsection{Structural transition from $N=40$ to $42$}

\begin{figure}
  \begin{center}
    \resizebox{0.47\textwidth}{!}{
      \includegraphics*{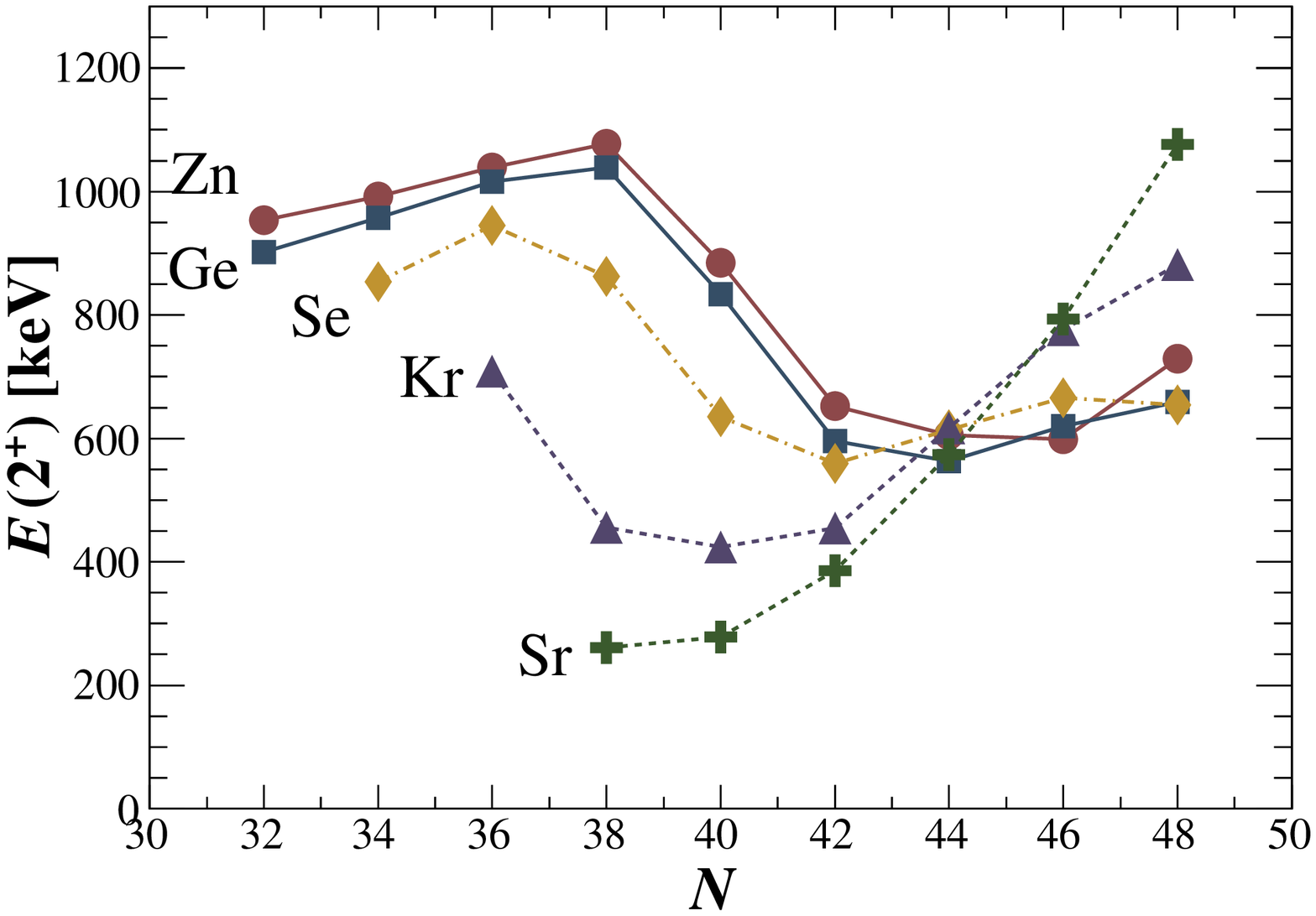}
    }%
    \caption{\label{fig:1} (Color online) Energy of the first $2^+$ state
      of Zn (filled circles), Ge (squares), Se (diamonds), Kr (triangles)
      and Sr (crosses) nuclei.}
    \vspace{0.5cm}
    \resizebox{0.47\textwidth}{!}{
      \includegraphics*{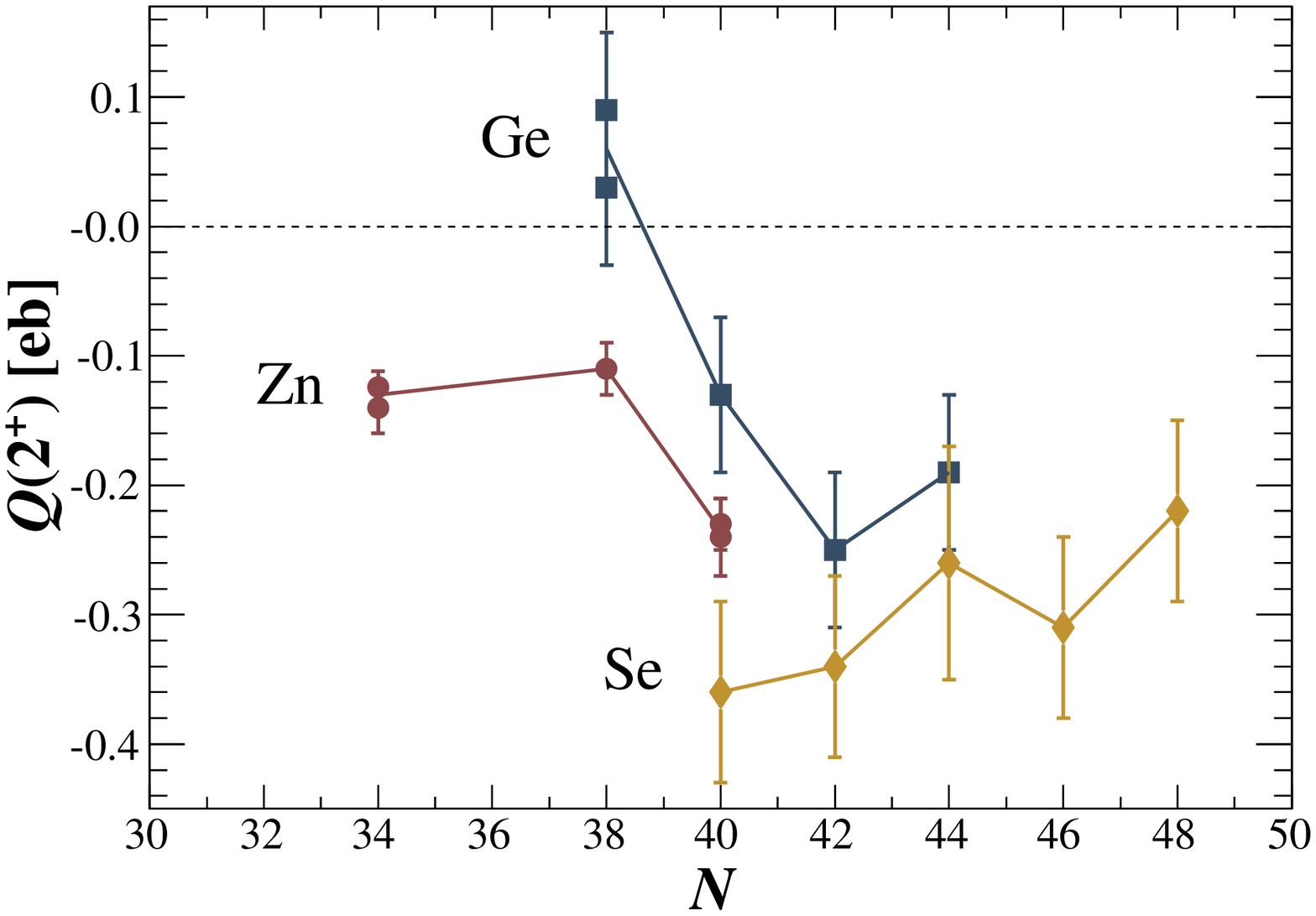}
    }%
    \caption{\label{fig:Q} (Color online) Measured $Q(2^+_1)$ (eb)
      values for Zn (filled circles), Ge (squares), Se (diamonds), Kr
      (triangles) and Sr (crosses) nuclei~\cite{sto}.}
    \vspace{0.5cm}
    \resizebox{0.47\textwidth}{!}{
      \includegraphics*{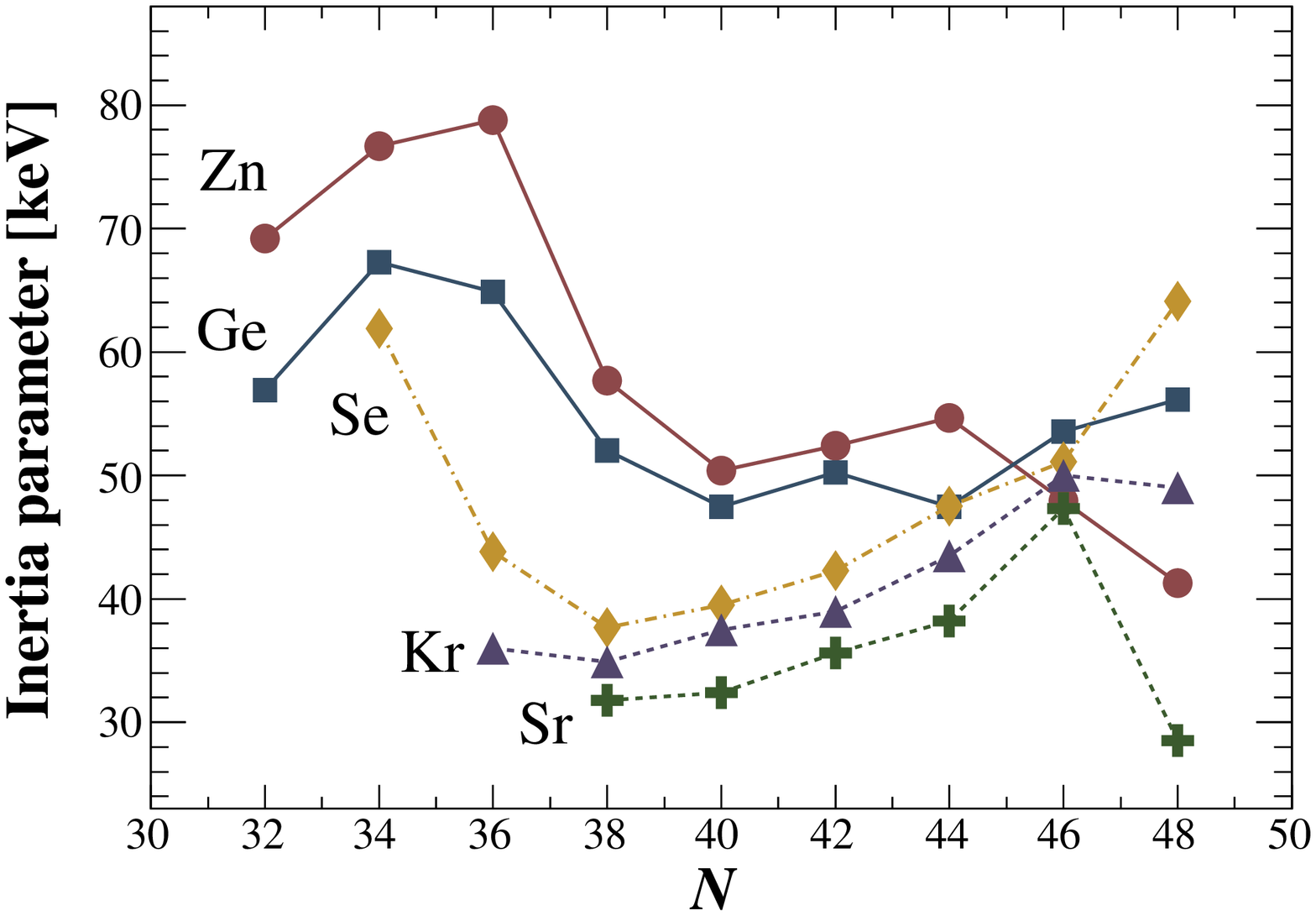}
    }%
    \caption{\label{fig:2} (Color online) Parameter of inertia
      $A=\hbar^2/2\Im$ from the $6^+\rightarrow4^+$ experimental
      energies for Zn (filled circles), Ge (squares), Se (diamonds),
      Kr (triangles) and Sr (crosses) nuclei ($6^+$ energies for
      $^{74,76}$Zn were taken from an unpublished work~\cite{fau}).}
  \end{center}
\end{figure}

In a first attempt to understand the development of collectivity in
the even-even Zn chain and to help in determining its nature, it is
interesting to consider the Zn data within the regional
systematics. The energies of the first $2^+$ states \textit{vs}
neutron number for $30\le Z\le 38$ even-even nuclei are shown in
Fig.~\ref{fig:1}. Two groups with different behavior can be
distinguished at once: the $E(2^+_1)$ is minimum for Sr and Kr close
or at $N=38$, while there is a maximum for Zn and Ge for exactly the
same number of neutrons. As already pointed out some time
ago~\cite{ham} this region provides a nice illustration of the
concepts of reinforcing and switching of shell gaps: the second group
(\textquotedblleft S-group\textquotedblright) appears indeed to be
dominated by $N=38$--$40$ fragile spherical gaps, reinforced by the
proximity of the $Z=28$ strong spherical gap, while the first
(\textquotedblleft D-group\textquotedblright) appears to be dominated
by the mutual reinforcement of $Z,N=38$ deformed gaps. The Se chain
would correspond to some transition between the two
regimes. Fig.~\ref{fig:Q} shows the measured $Q(2^+_1)$ values as a
function of neutron number for Zn, Ge and Se: globally, some
transition is observed at $N\simeq40$, from smaller absolute values
(including one compatible with $0$), to larger absolute values that
are, however, all negative. At $N=40$, the value for Zn gets closer to
those for Se, characteristic of a more prolate \textit{average}
intrinsic shape. We note that no measured $Q(2^+_1)$ is compatible
with intrinsic oblate shape for $N\ge 40$ for any of the three
isotopic chains.

\begin{figure}
  \begin{center}
    \resizebox{0.47\textwidth}{!}{
      \includegraphics*{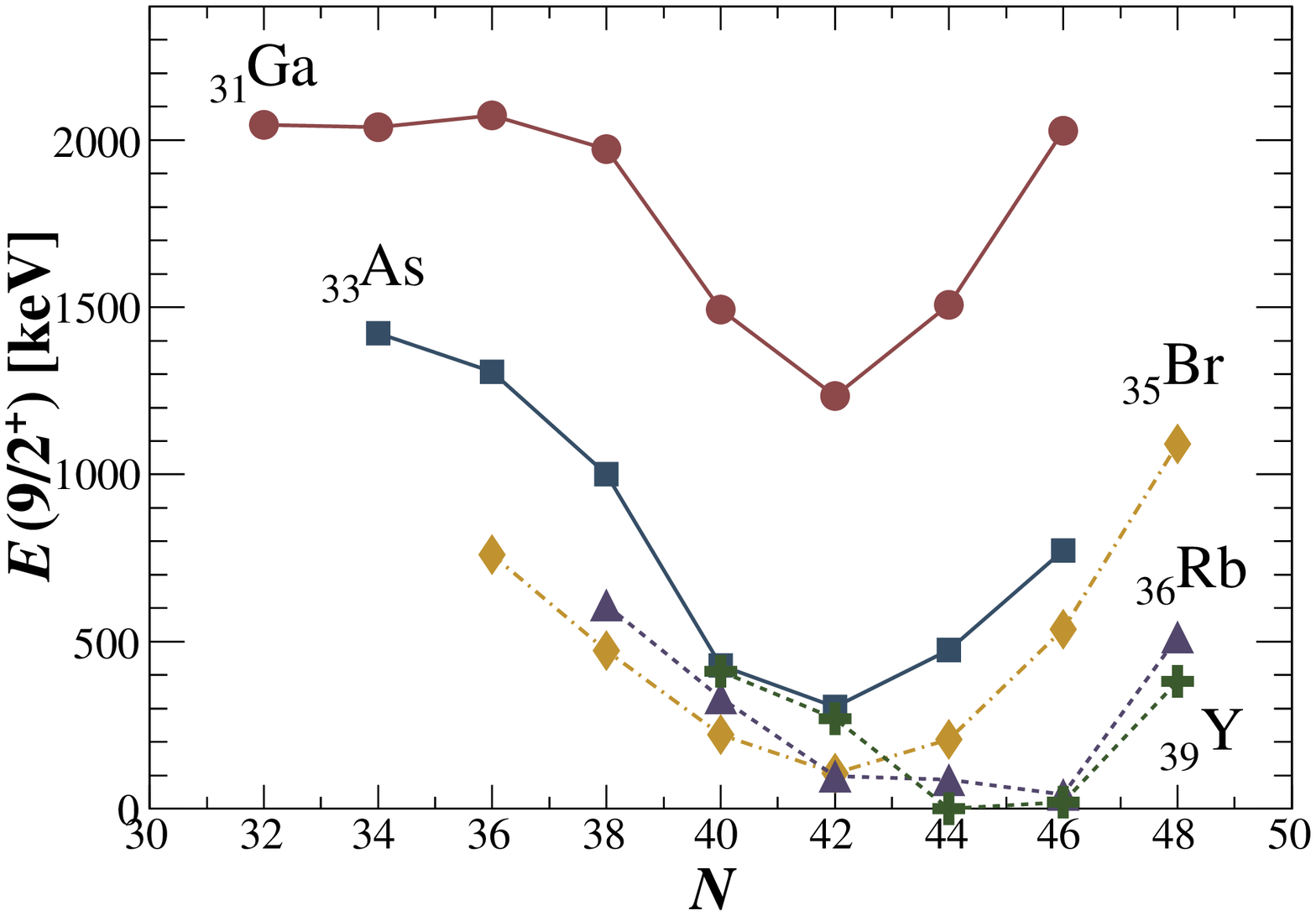}
    }%
    \caption{\label{fig:g92} (Color online) Energy of the first $9/2^+$
      state in the odd-proton Ga (filled circles), As (squares), Br
      (diamonds), Rb (triangles) and Y (crosses) nuclei.}
    \vspace{0.5cm}
    \resizebox{0.47\textwidth}{!}{
      \includegraphics*{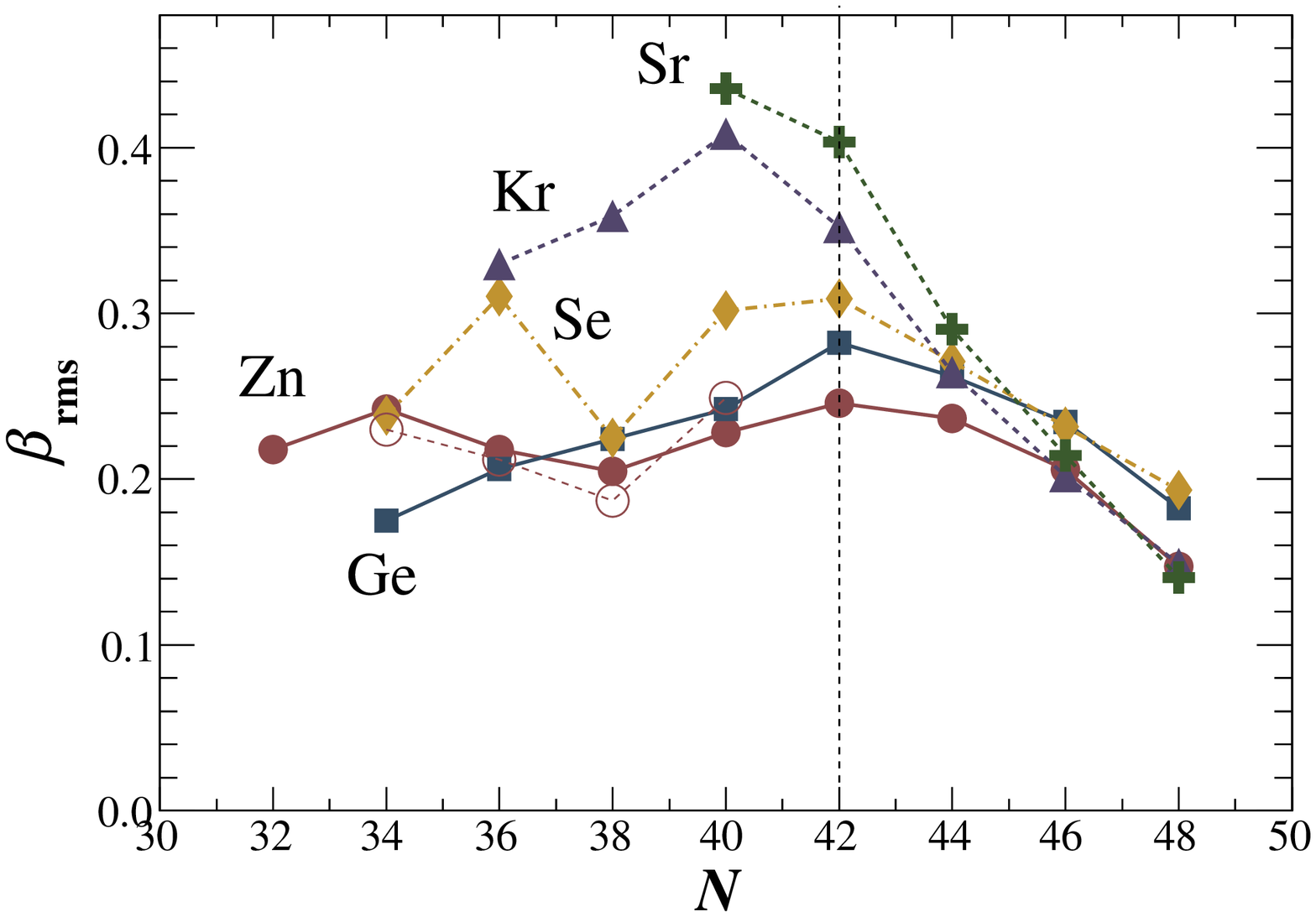}
    }%
    \caption{\label{fig:beta} (Color online) Root-mean-square~(rms)
      deformation parameters for Zn (filled circles), Ge (squares), Se
      (diamonds), Kr (triangles) and Sr (crosses) nuclei (no error
      bars are reported here due to the intrinsic uncertainty of
      Eq.~\ref{eq:brms}). $\beta_2$ experimental values for
      Zn~\cite{neu} are represented by open circles.}
  \end{center}
\end{figure}

It is interesting to consider also the inertia parameters
$A=\hbar^2/2\Im$ corresponding to the $6_1^+-4_1^+$ energy differences
(which allows to get rid of band mixing effects to some extent). These
values are reported in Fig.~\ref{fig:2} as a function of the neutron
number. The largest value for the moment of inertia is found at $N=38$
for $Z=34$ and $Z=36$. By contrast, for both $Z=30$ and 32 the moment
of inertia increases up to $N=40$ and then is somewhat stabilized with
values of the inertia parameter $A= \hbar^2/2\Im$ scattered around 50
keV hinting at the apparent stabilization of certain dynamic regime
from $N=40$ and all along the $\nu 1g_{9/2}$ filling.

 An additional piece of information which documents a maximum of
 quadrupole coherence at $N=42$ is provided by the energy systematics
 of the first $9/2^+$ states \textit{vs} neutron number for $31\le
 Z\le 39$ odd-even nuclei as shown in Fig.~\ref{fig:g92}. These states
 are interpreted, on the oblate side, as the head of the band built on
 the rapidly down-sloping $9/2^+[404]$ Nilsson orbital stemming from
 the proton $1g_{9/2}$ spherical orbit, or alternately, on the
 slightly prolate side, as the favored members of a rotation aligned
 band built on the $1/2^+[440]$ Nilsson orbital stemming from the same
 shell. One sees that these opposite parity states have a minimum in
 energy at $N=42$ for Ga (Zn$+ 1$ proton), As (Ge$+ 1$ proton) and Br
 (Se$+ 1$ proton).

Finally, the new values of the reduced transition probability $B(E2;
2^+_1\rightarrow0^+)$ for $^{72,74}$Zn obtained from the present work,
when inserted in the existing systematics (see Fig.~\ref{be2sys})
allow to confirm the same interesting feature: the maximum value for
Zn, Ge and Se is reached systematically at $N=42$. As is well known,
the most probable deformation parameter can be evaluated by:
\begin{equation}\label{eq:brms}
  \beta_{\mathrm{rms}} \gtrsim \frac{4\pi\sqrt{5}}{3ZR^2_0}\sqrt{B(E2; 2^+_1\rightarrow 0^+)} 
\end{equation}
with $R_0=1.2 A^{1/3}$~fm. This relation which was originally derived
by Kumar~\cite{kum} is model independent in the sense that no
intrinsic shape is assumed; the only assumption is a uniform charge
distribution inside a sharp oscillating surface. This quantity is a
measure of the collectivity in general: large values of
$\beta_{\mathrm{rms}}$ should be understood as an indication of large
deformation or big vibration amplitudes or both. Only in the case of a
permanent axially deformed intrinsic shape all collectivity amounts to
the rotation of the shape and $\beta_{\mathrm{rms}}\equiv \beta_2
$. An application of this relation was made by Nolte \textit {et al.}
for this mass region~\cite{nol}, we present the updated systematics in
Fig.~\ref{fig:beta}. The $\beta_2$ parameter has been deduced in an
independent way from inelastic electron-scattering cross-section
measurements for $^{64-70}$Zn~\cite{neu}, and the values are also
reported in Fig.~\ref{fig:beta}. A nice agreement is found both in order
of magnitude and tendency \textit{vs} neutron number. From
Fig.~\ref{fig:beta}, a maximum of collectivity at $N=42$ for Zn, Ge and
Se is confirmed. The question is now: which kind of collectivity?

\begin{figure*}
  \resizebox{0.7\textwidth}{!}{
    \includegraphics*{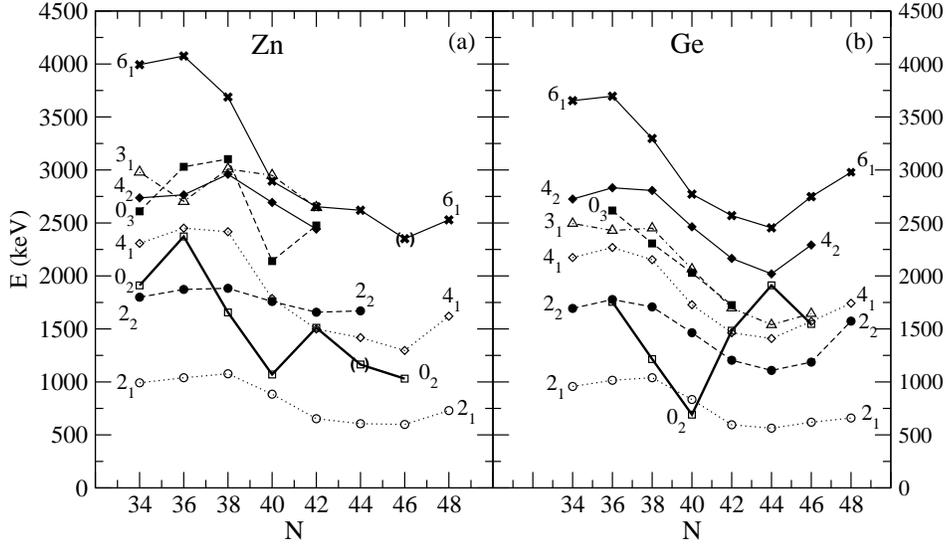}
  }%
  \caption{\label{fig:4} Systematics of the low-lying positive parity
    states in $^{64-78}$Zn and $^{66-80}$Ge. $6^+$ energies for
    $^{74,76}$Zn and $^{76}$Ge were taken from an unpublished
    work~\cite{fau}.}
\end{figure*}
\begin{figure}
  \resizebox{0.5\textwidth}{!}{
    \includegraphics*{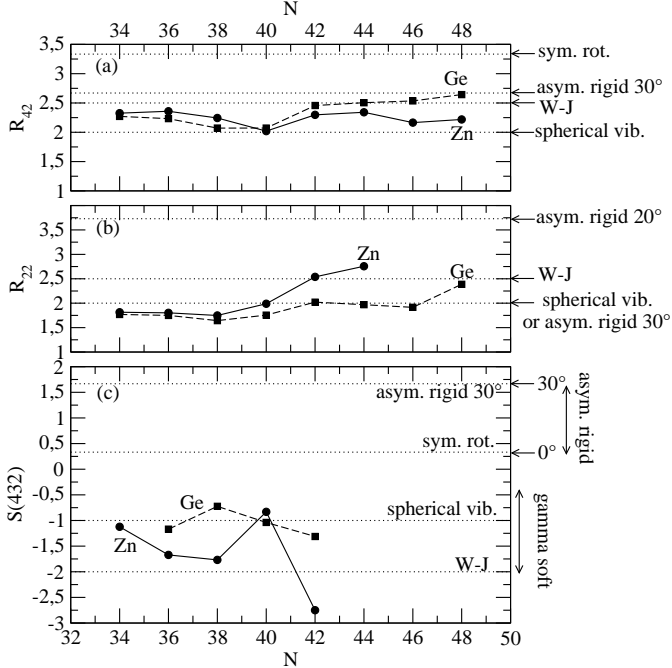}
  }%
  \caption{\label{fig:5} Systematics of the $R_{42}=E(4^+_1)/E(2^+_1)$
    and $R_{22}=E(2^+_2)/E(2^+_1)$ ratios and $S(432)$ signatures for
    Zn and Ge nuclei. On the right, the limits for the different
    geometrical models are indicated.}
\end{figure}
To answer this question a closer inspection of the energy systematics
of the low-lying levels of the Zn isotopes \textit{and} a comparison
with Ge is necessary. The energy evolution as a function of neutron
number for the $0^+_{2,3}$, $2^+_{1,2}$, $3^+_{1}$, $4^+_{1,2}$ and
$6^+_1$ states (when identified) in Zn and Ge is drawn in
Figs.~\ref{fig:4}(a) and (b). The trends of the structures of
even-even Zn and Ge nuclei show remarkable resemblance, especially the
$2^+_1$ and $4^+_1$ sequences. This goes along with our previous
remarks that these two isotope series are the best representatives of
the \textquotedblleft S-group\textquotedblright. Figs.~\ref{fig:5}(a)
and (b) show the $R_{42}=E(4^+_1)/E(2^+_1)$ and
$R_{22}=E(2^+_2)/E(2^+_1)$ ratios. The $R_{42}$ values lie in between
the values $2$ for a vibrator and $2.67$ for an asymmetric rotor
(Davydov-Filippov) with $\gamma=30^{\circ}$, and several are
compatible with the completely $\gamma$-soft (Wilets-Jean~(W-J) or
O(6) in IBM representation) limit of 2.5. At any rate, all remain far
away from the rotor value $3.33$. $R_{22}$ always keeps modest values
hinting at $\gamma_{\mathrm{rms}}\simeq 30^\circ$ and
$\gamma$-softness rather than developed axially asymmetric shape. A
similar conclusion was reached in Ref.~\cite{vdw} from the
consideration of the $R_{42}$ and $B(E2;4^+\rightarrow
2^+)/B(E2;2^+\rightarrow 0^+)$ systematics. In addition, it is
interesting to note that the closest values to the spherical vibrator
limit are found for both $R_{42}$ and $R_{22}$ for both Zn and Ge at
$N=40$. From $N=40$ to 42, either one of the two values (or both)
deviates immediately from this limit, which is never approached again
for $N\ge42$.  Finally, Fig.~\ref{fig:5}(c) shows the signature
$S(432)$ as defined by Zamfir and Casten~\cite{zam}:
\begin{equation}
  S(432)=\frac{\left[ E(4^+_\gamma) - E(3^+_\gamma)\right] -\left[
      E(3^+_\gamma) - E(2^+_\gamma)\right]}{E(2^+_1)}
\end{equation}
as a means to distinguish between $\gamma$-unstable and triaxial rotor
and a way to quantify $\gamma_{\mathrm{rms}}$. For the Zn even-even
isotopes, these values can be interpreted as $\gamma$-softness with
$\gamma_{\mathrm{rms}}\simeq15-20^\circ$ for $N\le38$, quasi harmonic
vibrator at $N=40$, and complete $\gamma$-softness at $N=42$
($\gamma_{\mathrm{rms}}= 30^\circ$). The complete $\gamma$-softness in
$^{72}$Zn is confirmed by the inspection of the level scheme: there is
a clear clustering of the $(2^+_2,4^+_1)$ states and
$(0^+_2,3^+_1,4^+_2,6^+_1)$ states close to the expected energies of
$2.5\cdot E(2^+_1)$ and $4.5\cdot E(2^+_1)$, respectively. In the case
of the latter cluster of states, one should keep the $0^+_2$ state
observed at 1499 keV out of it, otherwise this could lead to the fake
picture of a harmonic vibrator. It has been shown indeed that the
$0^+_2$ state in Ge is characterized by marked different structure
(sometimes quoted as \textquotedblleft intruder\textquotedblright)
with respect to other low-lying levels: this has been abundantly
discussed in the past and up to recently~\cite{
  ver2,car,for,kot,cho,toh,toh2,sug} (and references therein). Strong
similitudes are observed between the Zn and Ge chains in the global
trends of this $0^+_2$ state for both energy evolution
(Fig.~\ref{fig:4}) and reaction cross-section ratios $\sigma_{0^+_2}
/\sigma_{0^+_{gs}}$~\cite{ver2,jab} which hint at a similar situation
of coexisting structures in Zn. For instance, the measured $E0$
transition probability for the $0^+_2\rightarrow 0^+_{gs}$ transition
in $^{70}$Zn ($N=40$) is absolutely not compatible with the
vibrational picture~\cite{rei}. In $^{70}$Zn, the $0^+_3$ state is
visibly pushed away by the interaction with this \textquotedblleft
intruder\textquotedblright state from an otherwise perfectly clustered
2-phonon triplet of states that one would naturally expect from all
signatures $R_{42}$, $R_{22}$, and $S(432)$ as discussed above. We
note in passing that the non zero measured $Q(2^+_1)$ value for
$^{70}$Zn (see Fig.~\ref{fig:Q}) is \textit{not} in contradiction with
the harmonic picture since even in a nucleus whose equilibrium shape
is spherical, the cubic term can lead to a pseudo-rotational value for
the static quadrupole moment $Q(2^+_1)$.

In conclusion, the $B(E2)$ values for $^{72,74}$Zn obtained in the
present work introduce a change in the global trend of the Zn $B(E2)$
systematics which brings new support to the idea of a maximum of
collectivity at $N=42$ for the S-group in the $f_{5/2}pg_{9/2}$
shell. All experimental evidence points towards a transition from a
spherical oscillator at $N=40$ to complete $\gamma$-softness at
$N=42$. Only the missing knowledge of any $B(E2)$ ratios prevents us
from designating definitely $^{72}$Zn as a textbook example of
Wilets-Jean (or O(6) in IBM language) $\gamma$-soft nucleus. In the
complete $\gamma$-unstable picture, $Q(2^+_1)$ is identically zero. At
$N=44$, $\gamma$-softness apparently persists (from $R_{42}$, $R_{22}$
of Fig.~\ref{fig:5} and $B(E2)$ ratios as shown in Fig.~11 of
Ref.~\cite{vdw}) which means that $Q(2^+_1)$ would remain close to
zero. In view of the present discussion, the assumption of
$Q(2^+_1)_{{}^{74}\mathrm{Zn}}=0$ made in Ref.~\cite{vdw} appears as a
reasonable approximation. Our $\tau_{2^+_1}$ value for $^{74}$Zn is
then compatible with the Coulomb-excitation data of Ref.~\cite{vdw}
(see Fig.~9 in this reference), considering our error bars. Those data
appear to rule out completely any possible prolate intrinsic shape. In
addition, it is quite well and long established, mainly from the
systematic study of Ge isotopes~\cite{lec}, that there is no
experimental evidence of oblate intrinsic deformations beyond $N=40$
just above $Z=28$ (and no physical reason for their appearance). For
that reason, it is quite clear that the structure transition from
$N=40$ to 42 can in no way be associated with the appearance of a
rotational behavior connected with a permanently deformed intrinsic
shape.

\subsection{Possible microscopic origin of the collectivity development at $N=42$ close to $Z=28$}

\begin{figure}
  \begin{center}
    \resizebox{0.47\textwidth}{!}{
      \includegraphics*{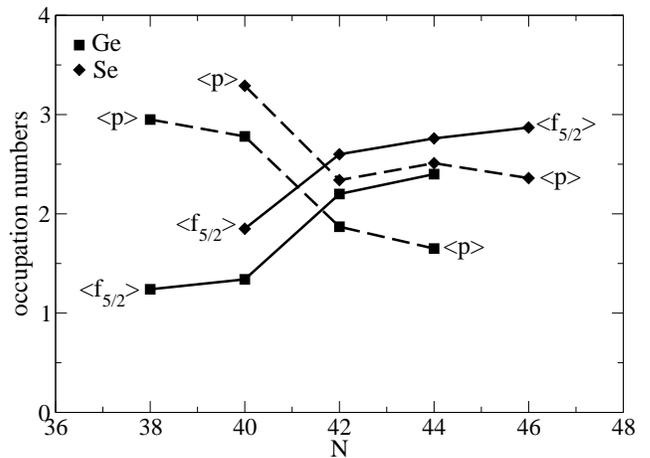}
    }%
    \caption{\label{fig:8} Measured proton occupation numbers in the
      Ge and Se isotopes from Ref.~\cite{rot}. Average global
      occupation numbers of the $2p_{3/2}$ and $2p_{1/2}$ orbitals
      (without distinction) $\langle p\rangle$ are connected with
      dashed lines and $\langle 1f_{5/2}\rangle$ with continuous
      lines.}
  \end{center}
\end{figure}

First clue to the possible microscopic origin of the $N=40$ to 42
transition can be looked for in the impressive body of data from
nucleon transfer experiments. Average proton occupations of the
$1f_{5/2}$, $2p$, $1g_{9/2}$ orbitals have been determined quite
precisely in the Ge and Se chains~\cite{ver2,car,rot} (see in
particular Table~2 in Ref.~\cite{ver2} and Fig.~7 and Table~7 in
Ref.~\cite{rot}). Those numbers are represented in a graphical way in
Fig.~\ref{fig:8} for the sake of convenience. There is a clear change
in the relative proton populations: from $N=40$ to 42 approximately
one proton is promoted on average from the $2p$ orbitals to
$1f_{5/2}$. Though such precise data are not yet available for the Zn
isotopic chain, one could imagine easily, due to the strong
similitudes in the structure of the nuclei of the \textquotedblleft
S-group\textquotedblright, that the situation is likely to be the same
also at $Z=30$. We see no trivial explanation to this change of proton
population: large-scale shell-model calculations show that proton
$1f_{5/2}$ and $1p_{3/2}$ effective single-particle energies do cross,
but at a higher number of neutrons (close to $N=46$, see Fig.~1 in
Ref.~\cite{SN}). That would mean that some correlation effects are at
play. For the sake of the present discussion, we just consider it as
an empirical fact. Then the simplest explanation of the observed
structure transition from $N=40$ to 42 might originate from a
Federman-Pittel~\cite{fed} like phenomenon. Since, as shown by all
recent large-scale shell-model calculations neutrons are already
present in the $1g_{9/2}$ orbital before the $N=40$ crossing, any
increase of the population of an orbit with large orbital angular
momentum, like $1f_{5/2}$, would trigger the deformation driving
proton-neutron interaction. $1g_{9/2}$ and $1f_{5/2}$ have equal
radial number and $\Delta \ell =1$. Furthermore, both have relatively
large orbital angular momenta. Hence, following the ideas developed by
Federman and Pittel, criteria for a strong proton-neutron interaction
effect are indeed met. This should be enough to qualitatively explain
the onset of collectivity (in the sense described above) from $N=40$
to 42. In Ref.~\cite{per} the strong proton core polarization beyond
$N=40$ was already ascribed to the attractive
$\pi1f_{5/2}$-$\nu1g_{9/2}$ monopole interaction.

In order to understand why the collectivity maximum at $N=42$ reveals
through $\gamma$-softness one should consider additional arguments. It
is well known that the presence of orbitals, nearby the Fermi level,
with high number of nodal planes parallel to the symmetry axis,
$n_\perp$, broadens the potential curve in the $\gamma$
coordinate. Here, $n_\perp=N-n_z$, where $N$ and $n_z$ are the
principal quantum number and number of nodes along the symmetry axis
in the $\left[ N n_z \Lambda\right]$ Nilsson labeling, respectively.
In the beginning of a deformation region, all orbitals have low
$n_\perp$ except those stemming from the lower shells, which, on the
contrary have maximum $n_\perp$. In our specific case, this
corresponds to the influence of the proton $7/2^-[303]$ orbital
stemming from the underlying $1f_{7/2}$ spherical shell. In
shell-model language, this simply means that the $\gamma$-softness
observed here is a natural result from the influence of the breaking
of the proton core.

\begin{figure}
  \begin{center}
    \includegraphics[width=.47\textwidth]{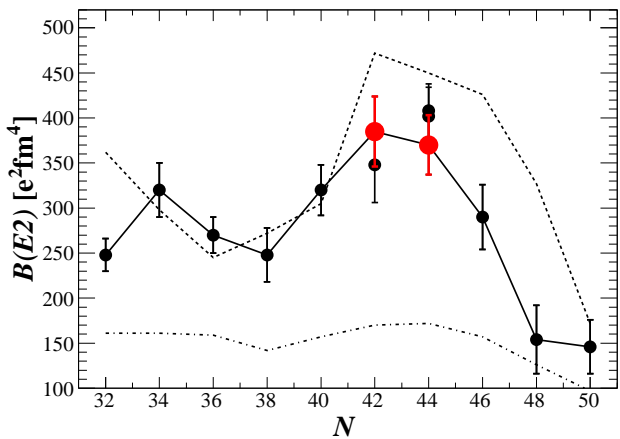}
    \caption{\label{znsys} (Color online) Experimental systematics of
      $B(E2)$ values for Zn isotopes compared to shell-model
      calculations using the LNPS interaction~\cite{sie} (dotted line)
      and the JUN45~\cite{hon} (dash-dotted line) with standard
      effective charges ($e_p=1.5e$ and $e_n=0.5e$). New values
      available from the present work are represented in red. The
      solid line is to guide the eye.}
  \end{center}
\end{figure}
Microscopic description of the $^{68}$Ni region has recently reached a
high degree of precision in the framework of the shell
model. Impressive agreement between experimental and calculated
magnetic moments of odd neutron-rich copper ground states has been
achieved~\cite{SN}. One of the major conclusions of Ref.~\cite{SN} is
that the only way to obtain such good agreement is by allowing proton
core excitations (promotion of proton from the $1f_{7/2}$ across the
$Z=28$ gap) in the calculations. An even larger valence space
($pf$-shell orbitals for protons and $f_{5/2}pg_{9/2}d_{5/2}$ orbitals
for neutrons), allowing also neutron excitations across the $N=50$
gap, has been used to account for the island of inversion phenomena in
neutron-rich Cr region~\cite{len}. A slightly revised version of the
LNPS interaction of Ref.~\cite{len} was produced by Sieja and
Nowacki~\cite{sie} which allows improved agreement with the
experimental $E(2^+)$, $B(E2)$ values and $g(2^+)$ factors in the Zn
isotopic chain~\cite{fio}. A transition from $N=40$ to $N=42$ is also
observed in the experimental $g$ factors of the $2^+_1$ states of
Zn. The results of those calculations for the $B(E2)$ values are
reported in Fig.~\ref{znsys}. While the agreement is perfect up to
$N=40$ the theoretical curve is systematically shifted upwards from
$N=42$ on. However, the maximum value, which should be, as deduced
from the present experiment, at $N=42$ is indeed
reproduced. Interestingly enough, the modification of the LNPS
interaction between Ref.~\cite{len} and Ref.~\cite{sie} decreases the
calculated $B(E2)$ value for the $N=42$ isotone $^{70}$Ni (degrading
the agreement with experiment), while it is already too large for
$^{72}$Zn as only two protons are added. In Fig.~\ref{znsys}, the
results from calculations performed in the valence space restricted to
$(1f_{5/2}2p_{3/2}1g_{9/2})$ using the JUN45 interaction~\cite{hon}
with standard effective charges ($e_p=1.5e$ and $e_n=0.5e$) are also
reported for comparison. It is quite clear that the extension to the
much larger valence space including also $\pi 1f_{7/2}$ and $\nu
2d_{5/2}$ represents a decisive improvement, being crucial for getting
the necessary degrees of freedom thus allowing the collectivity to
develop at the correct magnitude. However, we hope that this
discussion calls for a more careful balance of the collectivity which
is brought in by the inclusion in the valence space of the like
quadrupole partners $\nu g_{9/2}$-$\nu d_{5/2}$ with respect to the
influence of the proton-neutron $\pi f_{5/2}$-$\nu g_{9/2}$
interaction and to some extent, the influence of proton-core
excitations.

\section{Summary}
In summary, the lifetime measurement of the 2$^{+}_{1}$ states in
$^{72,74}$Zn has been performed at the LISE spectrometer at GANIL
using the RDDS method. The lifetime values were determined to be
17.9(18) and 27.0(24)~ps for $^{72,74}$Zn, respectively, which
correspond to $B(E2;2^{+}_{1}\rightarrow 0^{+})=385(39)$ and
$370(33)$~$e^2$fm$^4$.
The $B(E2)$ systematics obtained in the present work, when added to a
careful inspection of other experimental quantities available for the
neighboring nuclei, brings new support to the idea of a systematic
maximum of collectivity at $N=42$ for Zn, Ge and Se nuclei. In
addition, available signatures from the experimental spectra point
towards a transition from a spherical oscillator at $N=40$ to complete
$\gamma$-softness at $N=42$. From the whole body of data $Q(2^+_1)$
should be $0$ for $^{72}$Zn and close to it for $^{74}$Zn, a value
which is compatible with our data.

\begin{acknowledgments}
We are thankful to the GANIL staff for the technical help and to the
EXOGAM Collaboration for providing the segmented-clover detectors. We
also thank Dr.~K.~Sieja and Dr.~F.~Nowacki for providing us their
recent calculations prior to the publication in Ref.~\cite{sie}. This
work was partially supported by the German BMBF (under contract No.~06
DA 9040 I-1), the Polish Ministry of Science and Higher Education
(Grant No.~N~N202 309135 and N~N202 109936) and the Hungarian
Scientific Research Fund (OTKA, contract No.~K68801).
\end{acknowledgments}

\end{document}